\documentclass[12pt]{iopart}
\pdfoutput=1

\usepackage{iopams}

\expandafter\let\csname equation*\endcsname\relax
\expandafter\let\csname endequation*\endcsname\relax

\usepackage{amsmath}
\usepackage{amssymb}
\usepackage{amsfonts}

\usepackage[T1]{fontenc}
\usepackage[utf8]{inputenc}
\usepackage{microtype}

\usepackage{mathrsfs}
\usepackage{anyfontsize}

\usepackage[linktocpage,breaklinks,pdfusetitle]{hyperref}
\usepackage{cite}
\usepackage[all]{hypcap}
\usepackage[usenames,dvipsnames]{xcolor}

\usepackage{tensor}
\usepackage{bm}

\usepackage{graphicx}

\usepackage{tikz}

\ifdefined\myext
\usetikzlibrary{external}
\fi

\usepackage{orcidlink}
\usepackage{verbatim}

\hypersetup{colorlinks=true,
            citecolor=RoyalBlue,
            linkcolor=RoyalBlue,
            urlcolor=RoyalBlue}

\newcommand{\be}{\begin{equation}}
\newcommand{\ee}{\end{equation}}

\newcommand{\pd}{\partial}

\begin{document}

\title{Tidally-induced nonlinear resonances in EMRIs
with an analogue model}

\author{David~Bronicki
  \orcidlink{0000-0001-9382-572X},$^{1}$
  Alejandro C\'ardenas-Avenda\~no\
  \orcidlink{0000-0001-9528-1826},$^{2,3,4}$\\
  Leo C.\ Stein \orcidlink{0000-0001-7559-9597}$^{1}$
}

\address{$^{1}$~Department of Physics and Astronomy, The University of
  Mississippi, University, Mississippi 38677, USA}
\address{$^{2}$~Programa de Matem\'atica, Fundaci\'on Universitaria
  Konrad Lorenz, 110231 Bogot\'a, Colombia}
\address{$^{3}$~Illinois Center for Advanced Studies of the Universe, University of Illinois at Urbana-Champaign, Urbana, Illinois 61801, USA}
\address{$^{4}$~Department of Physics, Princeton University, Princeton, NJ, 08544, USA}

\hypersetup{pdfauthor={Bronicki et al.}}

\date{\today}


\begin{abstract}
One of the important targets for the future space-based gravitational wave observatory
LISA is extreme mass ratio inspirals (EMRIs), where long and accurate waveform modeling
is necessary for detection and characterization.
Modeling EMRI dynamics requires accounting for effects
such as the ones induced by an external tidal field,
which can break integrability at resonances and cause significant dephasing.
In this paper, we use a Newtonian analogue of a Kerr black hole to study the
effect of an external tidal field on the dynamics and the gravitational waveform. 
We have developed a numerical framework that takes advantage of the
integrability of the background system to evolve it with a symplectic
splitting integrator, and compute approximate gravitational waveforms
to estimate the timescale over which the perturbation affects the
dynamics.
Comparing this timescale with the characteristic time under radiation reaction at resonance, we introduce a tool for quantifying the regime in which tidal effects might be
included when modeling EMRI gravitational waves.
As an application of this framework, we perform a detailed analysis of
the dynamics at one resonance to show how different entry points into the resonance
in phase-space can produce substantially different dynamics,
and how one can estimate bounds for the parameter space where tidal effects may become dominant.
Such bounds will scale as $\varepsilon \gtrsim C \, q$, 
where $\varepsilon$ measures the strength of the external tidal field, $q$ is the mass ratio,
and $C$ is a number which depends on the resonance and the shape of the tide.
We demonstrate how to estimate $C$ using our framework for the 2:3 radial to polar frequency resonance in our model system.
This framework can serve as a proxy for proper modeling of the tidal perturbation in the fully relativistic case.
\end{abstract}


\maketitle

\section{Introduction}
\label{sec:intro}

Future space-based detectors, such as the Laser Interferometer Space Antenna (LISA)~\cite{Barausse:2020rsu},
will allow studies of the gravitational waves (GWs) emitted when a small compact object,
with mass $m$, falls into a supermassive one, with mass $M$, in an extreme mass-ratio inspiral (EMRIs),
i.e., satisfying $q=m/M\ll 1$. EMRIs emit GWs in wavelengths inaccessible to ground-based detectors,
such as the LIGO/Virgo/KAGRA detectors. They are expected to happen rarely in any one galaxy,
but nevertheless thought to occur within the lifetime ($\sim4$ years) of the LISA mission~\cite{Babak:2017tow,Vazquez-Aceves:2021xwl}.

Generic EMRI orbits, which to zeroth order are bound Kerr geodesics,
can be highly eccentric and inclined, and therefore the emitted GWs
are expected to encode a rich phenomenology~\cite{Drummond:2022xej}.
Geodesic motion in Kerr is integrable~\cite{Carter:1968rr}, so
a bound Kerr orbit lies on a phase-space torus characterized by three frequencies:
one each associated with the radial, polar, and axial motion~\cite{Schmidt:2002qk,Drasco:2003ky}.
This torus is ergodically filled for most trajectories, but not for the resonant ones,
i.e., when two frequencies form a co-prime low-integer ratio.

When a system is Liouville integrable~\cite{contopoulos_order_2002},
i.e., there are the same number of degrees of freedom as independent
Poisson-commuting integrals of motion (such as geodesics in Kerr), the
dynamics around resonances does not show a distinctive character.
However, in the presence of a generic perturbation,
there can be qualitative changes to phase space. 
From the KAM theorem~\cite{arnol1978mathematical},
for sufficiently small perturbations,
almost all non-resonant tori are deformed but continue to foliate phase space.
Meanwhile, close to resonant tori, we expect nonlinear resonances and
chaos to develop.

Beyond the test-particle approximation, the dynamics deviate from
geodesic motion due to a force arising from the gravitational field
generated by the small particle. That perturbation is known as the
particle's self-force~\cite{Mino:1996nk,Quinn:1996am} and generates
dissipative (e.g., the radiation-reaction force) and conservative
effects (e.g., the advance of the pericenter angle in each
orbit).

The formative work on EMRIs was performed under the so-called ``adiabatic'' approximation~\cite{Hughes:2005qb}, which only captures the long timescale behavior~\cite{Ruangsri:2013hra}, in which radiation reaction effects are torus-averaged to compute the inspiral. The modern approach is to instead use the more general method of near-identity transformations~\cite{VanDeMeent:2018cgn}, which is able to also capture the oscillatory short-timescale dynamics. These methods may fail near resonances~\cite{Flanagan:2010cd}, where a new timescale is introduced, and a more general method of multiple timescales must be employed~\cite{Hinderer:2008dm}.
Near resonances, the effects of
perturbations are boosted (relative to their non-resonant size) by an
amount inversely proportional to the square root of the mass ratio, and cause
the system to evolve more rapidly, or ``jump,'' from one adiabatic orbit to
another~\cite{Flanagan:2010cd, Gupta:2021cno}.
While this effect is well explored~\cite{Flanagan:2010cd, vandeMeent:2013sza}
in current models~\cite{Hinderer:2008dm, VanDeMeent:2018cgn}, radiation reaction is not the only potential source of perturbation.

A ``jump'' can also arise from the conservative sector of the dynamics~\cite{LukesGerakopoulos:2010rc, Cardenas-Avendano:2018ocb, Lukes-Gerakopoulos:2021ybx,Destounis:2021mqv},
at the level of the geodesic of a non-integrable system. Within the adiabatic approximation, prolonged resonances have been shown to appear~\cite{LukesGerakopoulos:2010rc,Destounis:2021mqv}, that are expected to translate to ``glitches'' in the GW frequency~\cite{Destounis:2021mqv}.

Resonant effects can also be induced by the tidal field of nearby
stars or stellar-mass black holes, a dark matter distribution, or the
rest of the galactic
potential~\cite{Yang:2017aht,Bonga:2019ycj}. These tidal resonances
will cause a secular shift to the orbital angular momentum, that, if
properly modeled, would provide information about the distribution of
mass near a galactic-center black hole~\cite{Bonga:2019ycj}. Due to the intrinsic complexity and sources of resonant effects, a generic description is still lacking, and assessing the importance of resonant effects on future gravitational detections with LISA has proven difficult~\cite{Berry:2016bit,Speri:2021psr}.

In this work, we carry out an analysis
that highlights important features of the resonant phenomena that should be taken into account when modeling EMRIs. To this end,
we study a Newtonian Kerr
``analogue''~\cite{Will:2008ys,Glampedakis:2013dd,Eleni:2019wav} and add
a tidal perturbation. The analogue system is Euler's 18th century
problem of two fixed gravitating centers~\cite{Will:2008ys}, which is the unique stationary and axisymmetric Newtonian potential that shares several key properties of the Kerr metric~\cite{Glampedakis:2013dd}. In particular, it is integrable, bound orbits are characterized by three fundamental frequencies, and has the same recurrence relation for the mass multipole moments~\cite{Glampedakis:2013dd,Eleni:2019wav}.

This analogue is a compromise between a system complex enough to exhibit resonant effects, and still simple enough to be tractable. First we numerically evolve the trajectories of the perturbed system and compute approximate gravitational waveforms. Then we perform an analysis of dephasing times, and estimate a region of parameter space for which the characteristic timescale of the tidal perturbation at resonance is shorter than the characteristic timescale of radiation reaction at resonance. That is, we estimate a range of parameter space where tidal perturbations will need to be further considered in waveform modeling.  Similar to findings in previous work~\cite{Flanagan:2010cd, LukesGerakopoulos:2010rc}, we also found that the different entry points into the resonance in phase space can produce substantially different dynamics.

The remainder of this paper shows the details of the calculations that led to the above conclusions. In Sec.~\ref{sec:unpert} we give an overview of the Newtonian analogue, and in Sec.~\ref{sec:pert} we present how it is tidally perturbed. Our numerical scheme is presented in Sec.~\ref{sec:alg}. We characterize the resonant dynamics in Sec.~\ref{sec:kin}, before studying the approximate gravitational waves in Sec.~\ref{sec:gw}. Finally in Sec.~\ref{sec:disc-conc} we present our conclusions and perspectives. Throughout the paper we use geometric units in which $G_{\rm N}=c=1$, and set $M=1$ for all our numerical implementations.

\section{A Newtonian Analogue to the Kerr Spacetime}
\label{sec:unpert}

The system we study in this work is a particle moving in the oblate
version of Euler's potential
of two fixed centers~\cite{Will:2008ys}. In this section, we give a
brief summary of the system and only the equations directly required
for our purposes.  For a more complete description of the system, see Refs.~\cite{Will:2008ys,Glampedakis:2013dd,Eleni:2019wav}.

The gravitational potential is generated by two fixed point particles with some separation $2a$. To make the system analogous to
the Kerr spacetime, the masses of both particles are set to $M/2$
(making the potential parity symmetric) and an oblate characteristic
is enforced by setting the separation to an imaginary value
$2ia$~\cite{Vinti:1966vc, Will:2008ys}. Upon making the further choice
of aligning the separation with the $z$ axis, the potential can be expressed as
\be\label{eq:system-potential}
V=-\frac{M}{R^2\sqrt{2}}\sqrt{R^2+r^2-a^2}
\,,
\ee
where $R\equiv \sqrt{(r^2-a^2)^2+4a^2z^2} = r_a r_b$, and $r_a$ and $r_b$ are the distances from each fixed mass.

In the analogy to the Kerr spacetime, $M$ and $a$ are analogous to the mass and spin of the black hole, respectively, and also fully describe the solution. The analogue is endowed with a third independent commuting constant of motion, and thus, it is completely integrable~\cite{Will:2008ys}, as is the Kerr spacetime. Furthermore, the analogue has a multipolar expansion that follows exactly the same relation as the mass multipole moments of the Kerr solution~\cite{Will:2008ys, Glampedakis:2013dd,Eleni:2019wav}. A detailed list and analysis of the qualitative and quantitative similarities between the two problems can be found in Ref.~\cite{Eleni:2019wav}.

The motion can be separated in oblate spheroidal coordinates with
\begin{align}
    &\text{radial-like coordinate:} & r=a\xi&\in[0,\infty)\nonumber\\
    &\text{polar-like coordinate:} & \cos\theta=\eta&\in [-1,1]\nonumber\\
    &\text{azimuthal coordinate:} & \phi&\in [0, 2\pi)\label{eq:coord-ranges}
\end{align}
related to cylindrical coordinates by
\begin{align}
    \rho&=a\sqrt{1+\xi^2}\sqrt{1-\eta^2}\nonumber\\
    z&=a\xi\eta.\label{eq:coord-relations}
\end{align}

Since we are considering motion to be that of a test mass, we work
with quantities per unit test mass, i.e.,
$p=P/m$ and $H=E/m$, where $P$ and $E$ are the standard momenta and energy.
In particular, the momenta relate to the above coordinates by
\begin{align}
    p_\xi&=a^2\left(\frac{\xi^2+\eta^2}{1+\xi^2}\right)\dot\xi,\nonumber\\
    p_\eta&=a^2\left(\frac{\xi^2+\eta^2}{1-\eta^2}\right)\dot\eta,\nonumber\\
    p_\phi&=a^2(1+\xi^2)(1-\eta^2)\dot\phi,\label{eq:momenta}
    \intertext{and the Hamiltonian is} H&=\frac{p_\xi^2}{2a^2}\frac{1+\xi^2}{\xi^2+\eta^2}+\frac{p_\eta^2}{2a^2}\frac{1-\eta^2}{\xi^2+\eta^2}
    +\frac{p_\phi^2}{2a^2}\frac{1}{(1+\xi^2)(1-\eta^2)}-\frac{M\xi}{a(\xi^2+\eta^2)}.\label{eq:ham}
\end{align}
As two coordinates do not appear in the Hamiltonian, we have two immediate conserved quantities: from the time symmetry we have $H=$ const., and from the axial symmetry we have $\ell:=p_\phi=$ const. The Hamilton-Jacobi theory can be implemented to further separate the system and obtain the third and final independent conserved quantity:
\begin{align}
    \beta&=-2a^2H(1+\xi^2)+p_\xi^2(1+\xi^2)-\frac{\ell^2}{1+\xi^2}-2aM\xi\nonumber\\
    &=-2a^2H(1-\eta^2)-p_\eta^2(1-\eta^2)-\frac{\ell^2}{1-\eta^2}.\label{eq:third-const}
\end{align}
Following this separation, new expressions for the momenta arise:
\begin{align}
    p_\xi&=\pm\sqrt{\frac{2a^2H(1+\xi^2)+\frac{\ell^2}{1+\xi^2}+2aM\xi+\beta}{1+\xi^2}}\nonumber\\
    &=\frac{\pm 1}{1+\xi^2}\sqrt{P_\xi(H,\ell,\beta;\xi)},\nonumber
    \intertext{and}
    p_\eta&=\pm\sqrt{\frac{-2a^2H(1-\eta^2)-\frac{\ell^2}{1-\eta^2}-\beta}{1-\eta^2}}\nonumber\\
    &=\frac{\pm 1}{1-\eta^2}\sqrt{P_\eta(H,\ell,\beta;\eta)},\label{eq:momenta-functs}
\end{align}
where $P_\xi$ and $P_\eta$ are fourth order polynomials in $\xi$ and $\eta$, respectively.

In studying the motion of particles in bound orbits, the previous two equations become quite important. By requiring the arguments of the radicals to be positive (so as to have real-valued momenta), we see that the motion in $\xi$ and $\eta$ becomes constrained to be between the roots of $P_\xi$ and $P_\eta$ respectively. To sit above the separatrix (analogous to requiring we not sit in a plunging orbit), we assign $\xi_1=r_1/a$ and $\xi_2=r_2/a$ to be the largest and second largest roots of $P_\xi$.
The other two roots may be real or complex. For motion along the polar coordinate, we define $\eta_\text{max}$ to be the smallest positive root of $P_\eta$. $P_\eta$ is a biquadratic, so $-\eta_\text{max}$ is also a zero and is the lower turning point. The other roots of $P_\eta$ are also symmetric to one another and must both be real. The set of turning points $(r_1, r_2, \eta_\text{max})$ constitute an alternative set of constants of motion. To relate these constants to more traditional Keplerian style orbital constants, we define eccentricity $e$, semilatus rectum $p$, and inclination $I$ such that
\begin{align}
    r_1&=\frac{p}{1-e} \,, & 
    r_2&=\frac{p}{1+e} \,, &
    \sin I&=\eta_\text{max} \,.\label{eq:turning-points}
\end{align}

As in the Kerr solution, the radial and polar coordinates are not periodic functions of the proper time~\cite{Schmidt:2002qk}. However, when expressed in Mino time~\cite{Mino:2003yg}, the equations of motion separate further and periodicity becomes manifest. Following this insight, as found in Ref.~\cite{Eleni:2019wav}, the equations of motion become
\begin{align}
    \frac{d\xi}{d\lambda}&=\pm\sqrt{P_\xi(\xi)} &
    \frac{d\phi}{d\lambda}&=\ell\left[\frac{1}{1-\eta^2}-\frac{1}{1+\xi^2}\right]\nonumber\\
    \frac{d\eta}{d\lambda}&=\pm\sqrt{P_\eta(\eta)} &
    \frac{dt}{d\lambda}&=a^2(\xi^2+\eta^2).\label{eq:motion}
\end{align}
The last equation here is the defining relation of Mino-like time, $\lambda$.

We note that the fundamental frequencies of the system can be found
analytically. By considering the equations of motion directly, these
fundamental frequencies are derived in Appendix E of
Ref.~\cite{Eleni:2019wav}. We will make use of these frequencies in
Sec.~\ref{sec:gw}.

\section{Tidal Perturbation}
\label{sec:pert}

We now consider the effect of an additional gravitating body or a
congregate gravitational potential. This perturbing potential can for
example arise from a nearby supermassive black hole, a star orbiting the primary outside the orbit of the secondary, or an overall galactic potential due to nearby stars and dark matter. In general relativity,
the local effects of gravity -- such as an external tidal field -- can
be modeled by using Riemann normal coordinates~\cite{MTW}; the
dominant effect is captured by the electric part of the Riemann tensor,
\be
E_{ij}=R_{0i0j},
\ee
where $R_{\mu\nu\rho\sigma}$ is the Riemann curvature tensor.
Meanwhile in Newtonian gravity, we may Taylor expand an external
gravitational potential about the origin as
\be
V_\text{ext}=V_0 + V_j x^j + \frac{1}{2} V_{jk} x^j x^k + O(x^3)
\,.
\ee
The first term is a constant and is removed by redefining zero
energy. The second, linear, term results in a constant force on both
the central body and the test particle equally, and
can be removed by choosing a freely falling frame for the binary.
This leaves, to lowest order,
\be
V_\text{ext}=\frac{1}{2} V_{ij} x^i x^j
\,.\label{eq:pert-potential}
\ee
Although $V_{ij}$ may in general depend on time, we will consider no time dependence for simplicity.

Note that in our units, $H=E/m$ (and by extension $V$ and
$V_\text{ext}$) is dimensionless.
Since $x_i$ has dimensions of mass,
$V_{ij} = \pd_{i}\pd_{j} V_{\text{ext}}$ has units of $1/M^2$.  We will define
$\varepsilon$ and $A_{ij}$ such that $V_{ij}=\varepsilon A_{ij}/M^2$,
where $M$ is the mass of the central body, $A_{ij}$ is dimensionless
and order unity, and we capture the smallness of the external tidal
field by $\varepsilon\ll 1$.
Eq.~\eqref{eq:pert-potential} now becomes
\be
V_\text{ext}=\frac 1 2 \varepsilon A_{ij} \frac{x^i}{M}\frac{x^j}{M}
\,.\label{eq:pert-dimensionless}
\ee
If the tidal force is due to a gravitating body of mass $M_*$ at a
distance $d$ from the center of the system, then
\be
\varepsilon\sim \frac{M^2 M_*}{d^3}
\,.\label{eq:pert-amplitude}
\ee

As an illustrative case, consider the environment in the center of our own galaxy, where we have a BH with mass $M=M_{\text{Sgr\,A}^*}\sim 4\times 10^6 M_\odot$, expected to be surrounded by a population of stellar-mass black holes $M_*\sim40 M_\odot$ with a mean distance of $d\sim 7.5\times 10^{11}$ meters~\cite{Emami:2019uty}. With these values, the tidal strength is $\varepsilon\sim 5\times 10^{-12}$.
We can also consider a system in which the primary is a supermassive
black hole with $M=10^9\,M_{\odot}$, and is perturbed by another
supermassive black hole, with $M_{*}=10^6\,M_{\odot}$. By considering
the perturber to be near by with $d\sim 4\times 10^{13}$ meters,
e.g., during the late stages of the merger of galaxies,
we find a stronger perturbation amplitude of $\varepsilon\sim 5\times10^{-8}$.

As the detection rates and specifics of these different types of systems are subject to several uncertainties~\cite{Yang:2017aht,Barausse:2020rsu,Vazquez-Aceves:2021xwl}, for our analysis we consider cases with perturbation amplitudes in the range $\varepsilon\in [10^{-9}, 10^{-5}]$.  Furthermore, to reduce the parameters modeled, we make the arbitrary, symmetry-breaking choice of
\be
A=
\begin{bmatrix}
1.2 & 1.3 & 1.2\\
1.3 & 1.3 & 1.2\\
1.2 & 1.2 &-2.5
\end{bmatrix}\,.\label{eq:pert-matrix}
\ee
Our results are independent of the choice of these numbers, provided that the constraints of symmetric and trace free are satisfied, these number are of order unity, and that there is no accidental symmetry (such as axial symmetry).

\section{System Integration Algorithm}
\label{sec:alg}

Due to its desirable properties for long-duration integration, we implement a symplectic splitting integrator~\cite{Yoshida:1990zz} to evolve the system. As such, we split the Hamiltonian into
\be
H=H_0 + V_\text{ext},
\ee
where $H_0$ was discussed in Sec.~\ref{sec:unpert} and $V_{\text{ext}}$ was discussed in Sec.~\ref{sec:pert}. For a generic symplectic splitting method~\cite{Yoshida:1990zz},
the system state $(x, p)$ is evolved under $H$ by the scheme
\begin{align}
    \exp &(\Delta t\, D_H)(x, p)= \exp(a_1\Delta t D_{V_\text{ext}})\exp (b_1\Delta t D_{H_0})\cdots\nonumber\\
    &\exp(a_N\Delta t D_{V_\text{ext}})\exp (b_N\Delta t D_{H_0})\, (x, p) + \text{error}
    \,,\label{eq:symp-alg}
\end{align}
where the infinitesimal time evolution operator is defined by a Poisson bracket
\begin{align}
    D_H(A) &= \left\{A,H\right\},
    \label{eq:symp-evo-oper}
\end{align}
for a generic function $A:M\to \mathbf{R}$,
where $M$ is the phase space manifold.
We require the coefficients to be constrained by
\begin{align}
\label{eq:symp-coef-constraint}
    \sum a_i &= \sum b_i = 1,
\end{align}
so that over one step the system is evolved a total of $\Delta t$ through both $H_0$ and $V$. As with all numerical techniques, the order of accuracy depends on the number of stages ($N$) used and the coefficients ($a_i$ and $b_i$). Explicit systems of equations can be constructed from Eq. \eqref{eq:symp-alg} by either Taylor expansion of the exponentials or application of the Baker-Campbell-Hausdorff formula. We set $N=2$ and placed error constraints (discussed shortly) to arrive at
\begin{align}
    a_1&=1/4 & b_1&=2/3\nonumber\\
    a_2&=3/4 & b_2&=1/3.\label{eq:symp-coefs}
\end{align}
This choice produces an algorithm accurate to third order in time step and to first order in perturbation magnitude. That is, our single-time step (local) error terms are all of the form
\be
\text{error} = O (\varepsilon^i \Delta t^j) \,,
\ee
where either $i\ge 2$ or $j \ge 4$. Despite the relatively low order, we found $\Delta t = 0.3$ to retain sufficient long-term characteristics for this analysis, as is expected for symplectic algorithms in general.

Equation \eqref{eq:symp-alg} assumes exact or near exact methods exists to integrate each subsystem $H_0$ and $V_{\text{ext}}$. While $H_0$ can in principle be expressed analytically by relating the coordinates at time $t$ to those at time $t+b_i\Delta t$ with elliptic integrals, it is far simpler to numerically evolve the separated system in adapted coordinates. The precise numerical integrating method used to evolve each subsystem is not important; we chose to use the RK45 method of the \texttt{solve\_ivp} routine provided by \texttt{scipy}~\cite{2020SciPy-NMeth}.

This choice comes with a caveat, however. Since we intend to integrate
through many orbits, the system will pass through many turning
points. At each turning point, for instance at $\xi_2$, the relevant
equation of motion is locally
\be
\frac{d\xi}{d\lambda}\approx \pm A\sqrt{\xi - \xi_2}.
\ee
The issue here is twofold. First, the sign is ambiguous, and so care
would have to be taken to keep track of which direction we travel and
when this direction switches. Secondly, since the square root function
has unbounded derivative, this system violates the Lipschitz
condition for existence and uniqueness~\cite{MR2242407}.
We therefore implement new phase angle coordinates $\psi$ and $\chi$ to replace $\xi$ (or $r$) and
$\eta$, defined by
\begin{align}
    a\xi=r&=\frac{p}{1+e\cos\psi}\,,
    &
    \text{and} &&
    \eta&=\eta_\text{max}\cos\chi.\label{eq:phase-angles}
\end{align}
Unlike $\xi$ and $\eta$, $\psi$ and $\chi$ are monotonic, removing the
sign issue.  These definitions also analytically cancel the roots at the
turning points; after a bit of algebra, their time derivatives satisfy
the Lipschitz condition.  In particular, the new equations of motion are
\begin{align}
    \frac{d\psi}{d\lambda}&=\frac{\sqrt{-2a^2H\xi_1\xi_2(\xi^2+b\xi+c)}}{\xi}\nonumber\\
    \frac{d\chi}{d\lambda}&=\sqrt{-2a^2H(z_+-\eta^2)}\label{eq:phase-motion}
    \intertext{with}
    b&=\xi_1+\xi_2+\frac{1}{a H}\nonumber\\
    c&=(\xi_1+\xi_2)b-\xi_1\xi_2+\frac{\beta}{2a^2 H}+2
    \,,
    \label{eq:extra-coefs}
\end{align}
and $z_+$ is the square of the upper root of $P_\eta$.
Although the integration here is done with respect to Mino time $\lambda$, the integration is terminated once time $t$ has advanced by $b_{i}\Delta t$ so as to be consistent with the requirements of the symplectic integrator.

The integration under $V_{\text{ext}}$ is far simpler since it is only
a function of position and not momentum.  In Cartesian coordinates, we have
\begin{align}
  \dot{x}_i&=\left\{x_i , V_{\text{ext}} \right\}=0 \nonumber\\
  \dot{p}_i&=\left\{p_i , V_{\text{ext}} \right\}=-V_{ij}x^j.
\label{eq:unpert-motion}
\end{align}
So $x^i$ does not change, and $p_i$ changes linearly with time (a
``momentum kick'').  The new system state is then realized by
re-evaluating the various ``constants'' of motion.
Since this changes
the turning points, the phase angles $\chi$ and $\psi$ must also be
re-evaluated.

Summarizing, our algorithm consists of
  alternating between evolving under $H_0$ using a \texttt{scipy}
  integrator and evolving under $V_\text{ext}$ using the analytic
  ``momentum kick'' above, with coordinate changes interleaving the
  two techniques. A constant time step $\Delta t$ is inherent to our
  symplectic splitting scheme, while the time steps used internally by
  the \texttt{scipy} integrator are free to change. It is only
  important that the \texttt{scipy} integrator evolves a total of
  $b_i\Delta t$ for each stage and accumulates low global error (we
  set relative and absolute tolerances of $10^{-13}$).

\section{Dynamics at Resonance}
\label{sec:kin}

In this section, we delve into the dynamics at resonance when under a
tidal perturbation.  To this end, we will discuss three core
components: Poincaré sections, the KAM theorem, and resonance
angles~\cite{arnol1978mathematical}.

\subsection{Poincaré Sections}
\label{sec:kin:poincare}

\begin{figure}
  \centering
  \ifdefined\myext
    \tikzsetnextfilename{resonance-cartoon}
    \input{figures/resonance-cartoon}
  \else
    \def\relsstandalone{}
    \ifx\relsstandalone\undefined
    \input{resonance-cartoon}
    \else
      \includegraphics[width=0.6\columnwidth]{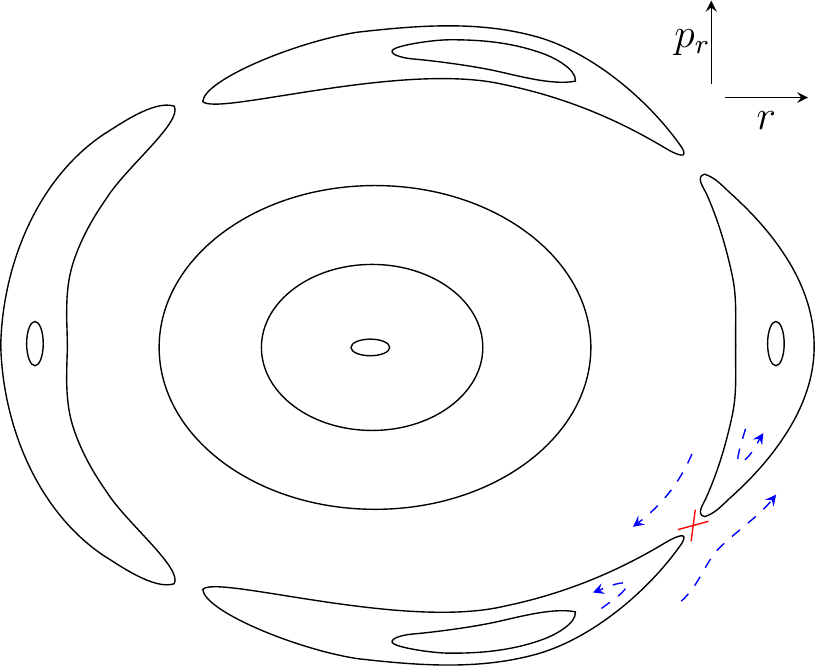}
    \fi
  \fi
  \caption{A schematic diagram showing the characteristics of a
    Poincaré surface of section for a perturbed system. The inner
    circles, where the original torus structure is preserved, are far
    from resonance, while the outer circles show the breaking of a
    resonant torus into a nonlinear resonance, forming a ``chain of
    islands.'' The red cross and blue flow lines show one (out of four) hyperbolic point and the direction the system evolves under the Poincaré map near it. The centers of the ``islands'' are elliptic
    points. With four elliptic points and four hyperbolic points, this
    diagram could correspond to a radial to polar frequency resonance
    of 1:4, 3:4, or 5:4, etc.}
  \label{fig:tori-break-cartoon}
\end{figure}

Let us briefly discuss a useful technique for graphically representing
the behavior of the dynamical system. The graphical representation,
known today as a Poincaré surface of sections or map of first recurrence~\cite{poincare1890probleme}, is obtained by considering the intersections of trajectories sharing the same value of energy with an arbitrary, but fixed, two-dimensional surface in phase space. Whenever the trajectory pierces the surface, the location in phase space is recorded, and the pattern generated on the surface constitutes the Poincaré surface of sections. The resulting figure, initially proposed by Poincaré, is extremely useful for systems of two degrees of freedom, as it is capable of providing a complete two-dimensional graphical representation of the four dimensional manifold, as sketched in Fig.~\ref{fig:tori-break-cartoon}. We describe its structure in the following section.

In our case, we choose the equatorial ($xy$) plane
as the surface that any bound orbit with a non-zero inclination will consistently intersect.
With the further constraints of accepting only ascending trajectories and $H={}$const.,
we are left with a four dimensional surface in phase space.
Thus, the two dimensional representation originally conceived by Poincaré cannot accurately capture the behavior of the perturbed system, as it will be a projection. However, as pointed out in Ref.~\cite{patsis1994using} if the fourth dimension is represented by a color variation (see Refs.~\cite{Zachilas:2012an,2014Richter,Lukes-Gerakopoulos:2016bup} for some examples), all the dimensions can be visualized, to some degree, on a colored 3D plot. An example of such 4D space of section is shown in Fig.~\ref{fig:4d}.

On the other hand, if we utilize the (approximate) exact azimuthal symmetry of the (perturbed) unperturbed case, we are able to further stipulate $\ell={}$const. and safely ignore the azimuthal coordinate. This reduces our visual representation to a traditional two dimensional surface, as seen in Fig.~\ref{fig:tori-break-cartoon} or in the $r-p_r$ plane of Fig.~\ref{fig:4d}.

\subsection{KAM Theorem}
\label{sec:kin:kam}

The Kolmogorov–Arnold–Moser (KAM) theorem~\cite{MR1656199} enters our
discussion as it makes statements regarding integrable systems when
perturbed. In particular, under a sufficiently small perturbation,
quasi-periodicity is retained for almost all orbits, resulting in
small deformations of the nested torus structure of integrable
Hamiltonian systems.  However, the theorem only guarantees this
behavior for ``sufficiently irrational'' ratios of fundamental
frequencies.  Our interest is exactly at the resonant tori, with
small-integer ratios of frequencies.  We find, as is generally the
case, that when sufficiently close to a resonance, the torus structure
breaks completely. This break of structure constitutes a topological
change, and can lead to both nonlinear resonances and
chaos~\cite{MR1656199}.

A cartoon of the structure of nonlinear resonances on a Poincaré
section is shown in Fig.~\ref{fig:tori-break-cartoon}.  The inner
sections of the figure are far from resonance and so the nested torus
structure is retained.  A resonant torus of the unperturbed system has
developed into a chain of ``islands'' of a single nonlinear
resonance.  Each island surrounds an elliptic fixed point, and they
are separated by an equal number of hyperbolic fixed points (red
crosses).
In the nomenclature of~\cite{vandeMeent:2013sza}, if an EMRI crossed
such a nonlinear resonance, it would be called a ``sustained''
resonance, since two or more fundamental frequencies have an
approximately constant ratio throughout the entire island.

The dynamics within the elliptic islands can be intuitively understood
by drawing an analogy with a nonlinear pendulum (the pendulum analogy
can be made even more precise in certain Solar system dynamics
systems~\cite{Murray1998}).  The pendulum's degree of freedom is
analogous to the slow oscillation of a Lissajous figure of the
nonlinearly resonant trajectory.  Approaching the elliptic fixed
point, the amplitude of the oscillation decreases.  The edge of the
island is analogous to the separatrix of the pendulum's motion ---
outside the island is the region where the pendulum rotates rather
than oscillates.  The analogy is strengthened in that the period of
oscillation becomes amplitude-independent (like a simple harmonic
oscillator) when approaching the elliptic fixed point, while the
period diverges when approaching the separatrix.

\begin{figure}
  \begin{center}
    \includegraphics[width=0.7\linewidth]{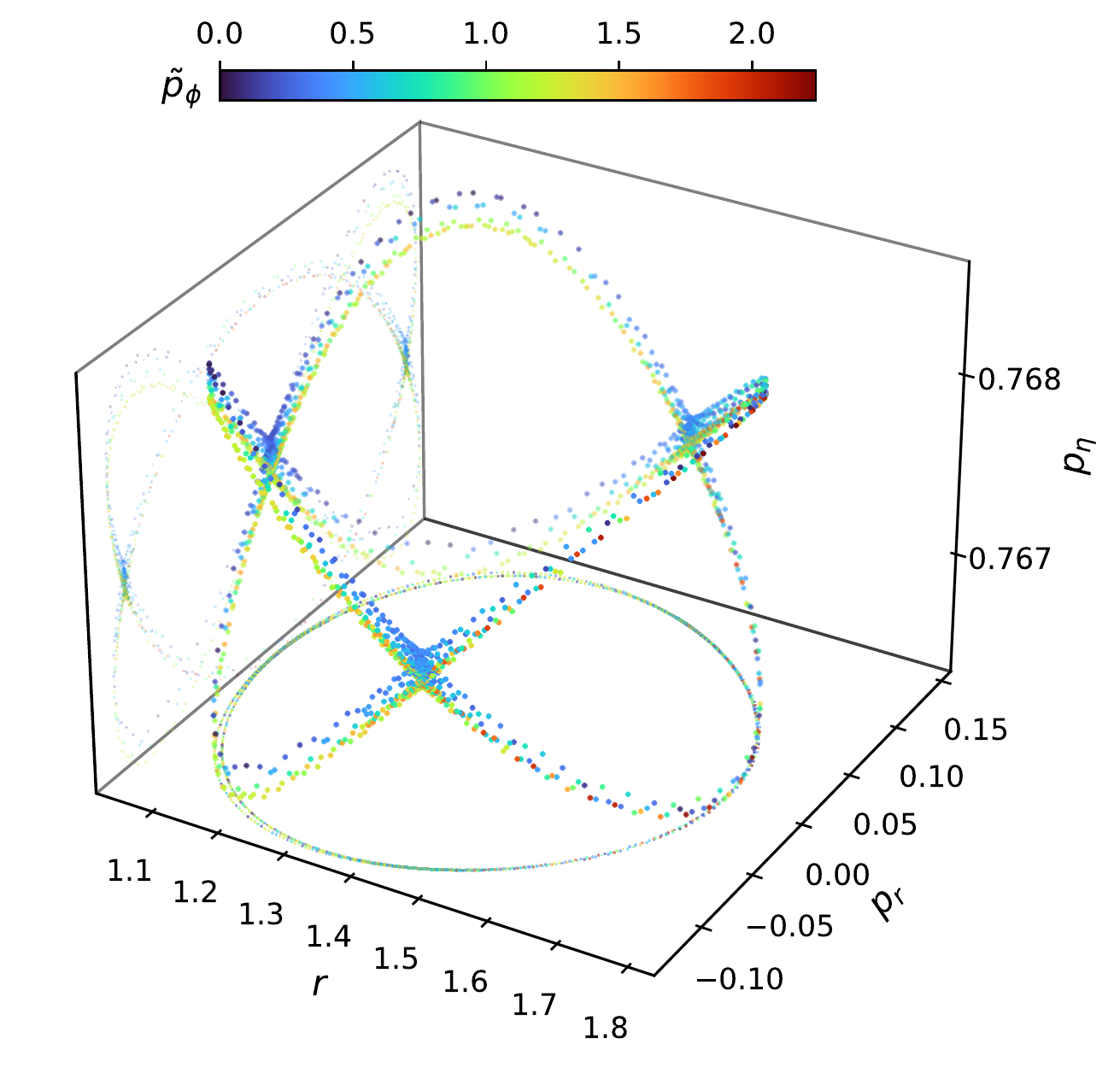}
  \end{center}
  \caption{%
      A 4-dimensional Poincaré section for a perturbed system with
      $\varepsilon=10^{-5}$, close to a nonlinear resonance.
      The colors correspond to the momentum associated with the
      coordinate $\phi$, which, for visualization purposes, has been
      scaled as $\tilde{p}_\phi=10^4(p_\phi-p_\phi^{\rm min})$.  The
      (minor) mixing of the colors indicates the presence of weakly
      chaotic orbits in the ``sea'' surrounding the quasiperiodic
      islands.  A single orbit in this sea can pass close to a
      hyperbolic fixed point, hence the sea can completely surround
      the elliptic islands.  The islands and surrounding sea are
      extremely narrow when projected onto the $(r,p_{r})$ plane, but
      it is apparent that the KAM curve has non-zero width.}
    \label{fig:4d}
\end{figure}

The loss (or non-existence) of a constant of motion can result in trajectories which project down to \emph{areas} in the $r-p_r$ plane (see e.g. Fig.~\ref{fig:4d}), despite being confined to a 3-torus within the 6-dimensional phase space. This makes it difficult to distinguish whether or not the torus structure has actually broken~\cite{Lukes-Gerakopoulos:2016bup}. A full 4-dimensional Poincaré section would distinguish the two scenarios; therefore we use a 3-dimensional projection with color encoding the fourth dimension~\cite{patsis1994using}. The mixing of
colors or the appearance of irregular behavior on the
three-dimensional projection signals the
breaking of a KAM torus. Thus, when studying the perturbed system, we
first use the 3D colored projection to check that the trajectory
belongs to a broken torus structure. This method lets us
assess the 4D smoothness of geometrical structures that appear in
the 3D projections, and thus find points near non-linear resonances. Then we return to the 2D projection, as if we only had two degrees of freedom, to study the dynamics more easily.

\subsection{Resonance Angles}
\label{sec:kin:res-angle}

\begin{figure}
  \begin{center}
    \includegraphics[width=0.7\linewidth]{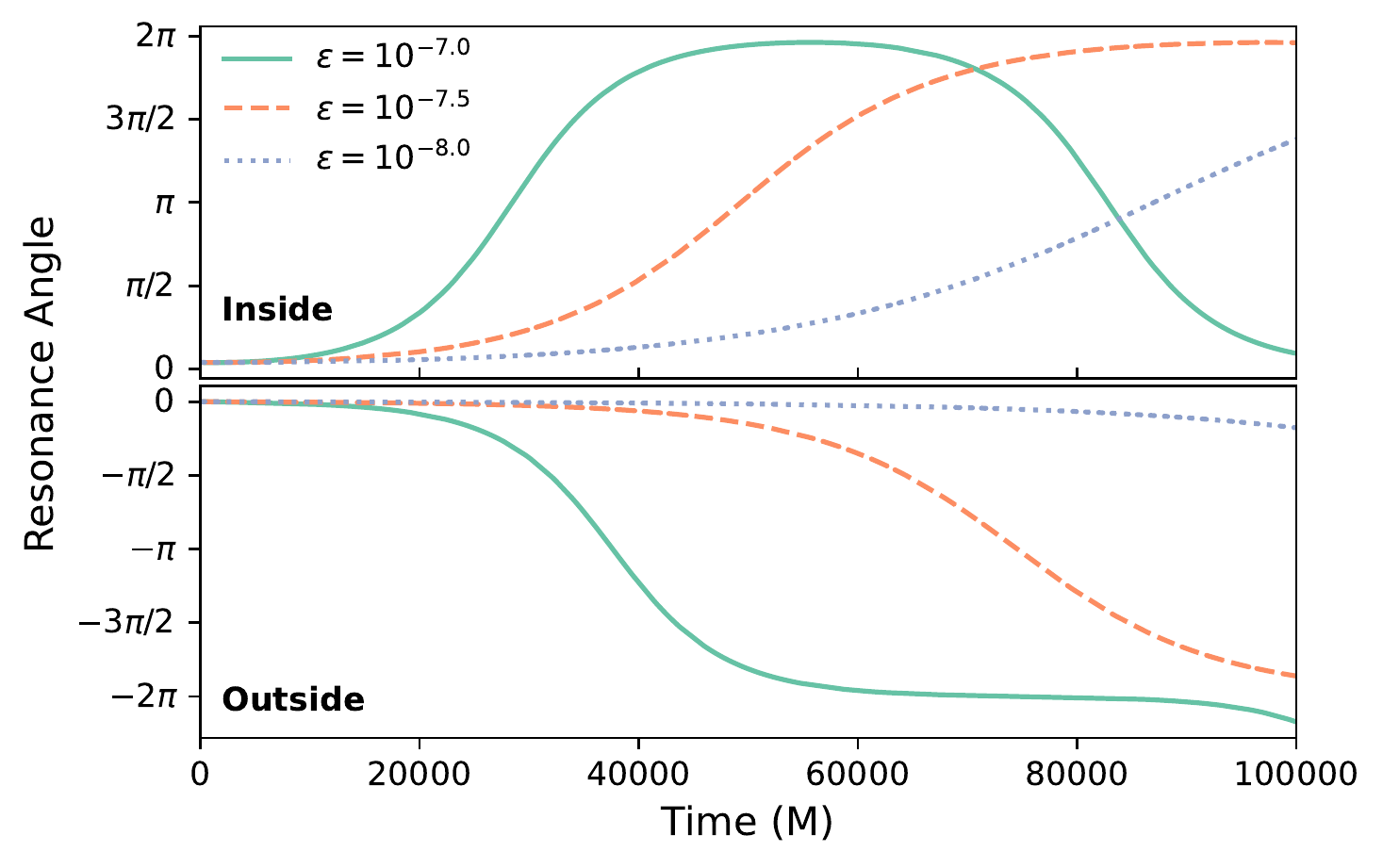}
  \end{center}
  \caption{%
      Dynamics of the resonance angle $\Psi$ inside the elliptic
      region and outside. Various levels of perturbation are shown for
      each plot. Inside the islands (upper panel) shows oscillatory
      motion, while the exterior (lower panel) shows rotating
      behavior, where $\Psi$ grows without bound.  We also note a
      shorter period of oscillation and faster rotation for larger
      perturbation amplitudes.}
    \label{fig:res-angle:pert-sweep}
\end{figure}

When considering (orbital) dynamics at resonance, one finds by definition that two of the system's phase angles evolve together with a specific rational ratio remaining approximately constant. As a result, one may define a ``resonance angle'' to be the difference of these phase angles with integer coefficients~\cite{Murray1998,Gair:2010iv,Seto:2013an}. We will focus on a system with a $2:3$ polar to radial frequency resonance, that is,
\begin{align}
    \psi&\approx 3\omega_0 t+\psi_0 &&\text{radial phase angle}\notag\\
    \chi&\approx 2\omega_0 t+\chi_0 &&\text{polar phase angle}\notag\\
    \Psi:=3\chi-2\psi&\approx \text{constant} &&\text{resonance angle},
\end{align}
where $\psi$ and $\chi$ were defined in Eq.~(\ref{eq:phase-angles}).
When a perturbation is applied, however, the phase angles will no longer evolve in such a simple fashion (it will now depend on a multi-periodic function), and so non-trivial dynamical information can be garnered from the resonance angle. In Fig.~\ref{fig:res-angle:pert-sweep}, we show how the resonance
angle evolves near resonance for various amplitudes of
perturbation. We can see a qualitative difference between trajectories inside the elliptic region to those outside. As expected from the pendulum model, we observe oscillating behavior only inside.

\begin{figure}
  \begin{center}
    \includegraphics[width=0.7\linewidth]{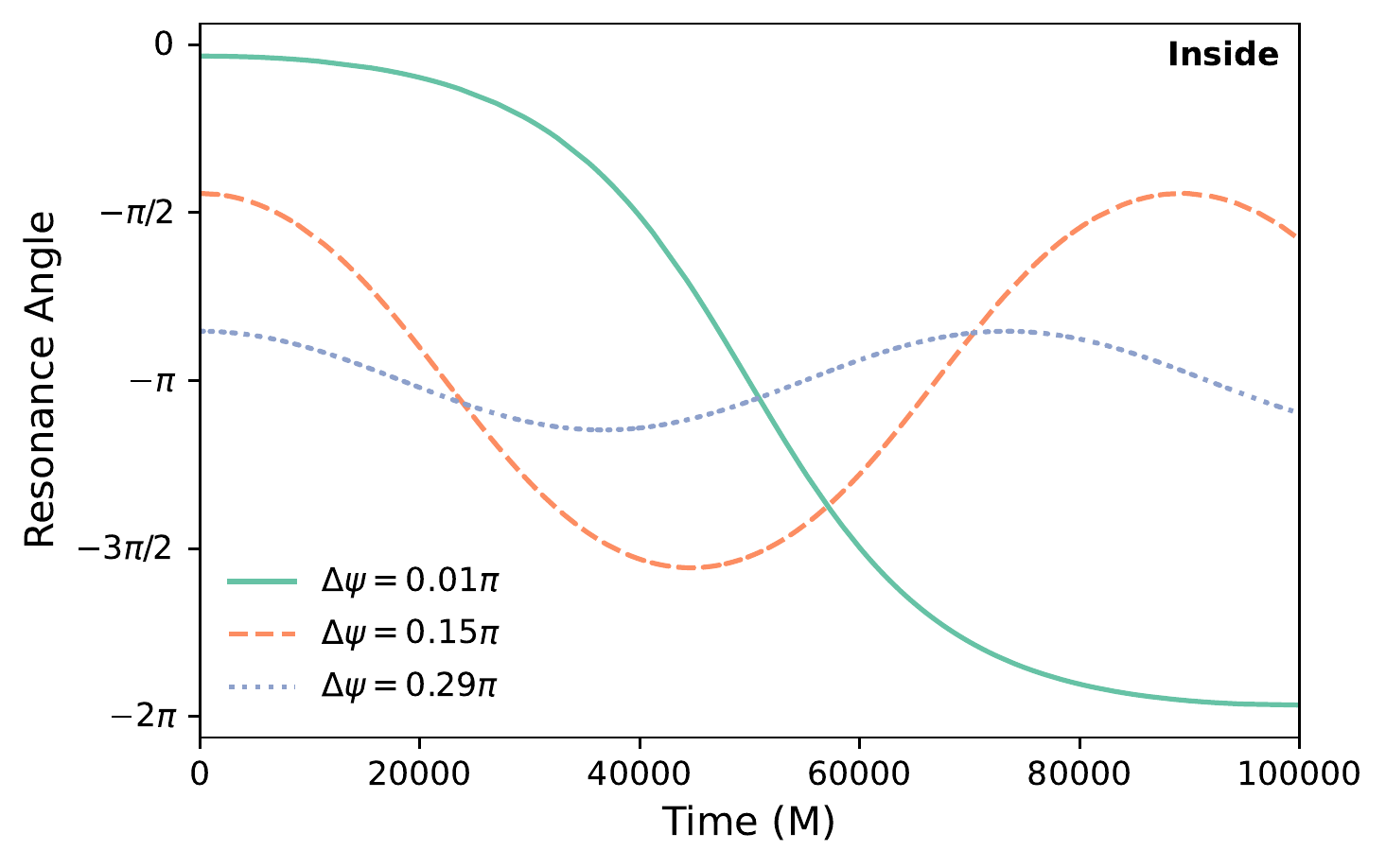}
    \includegraphics[width=0.7\linewidth]{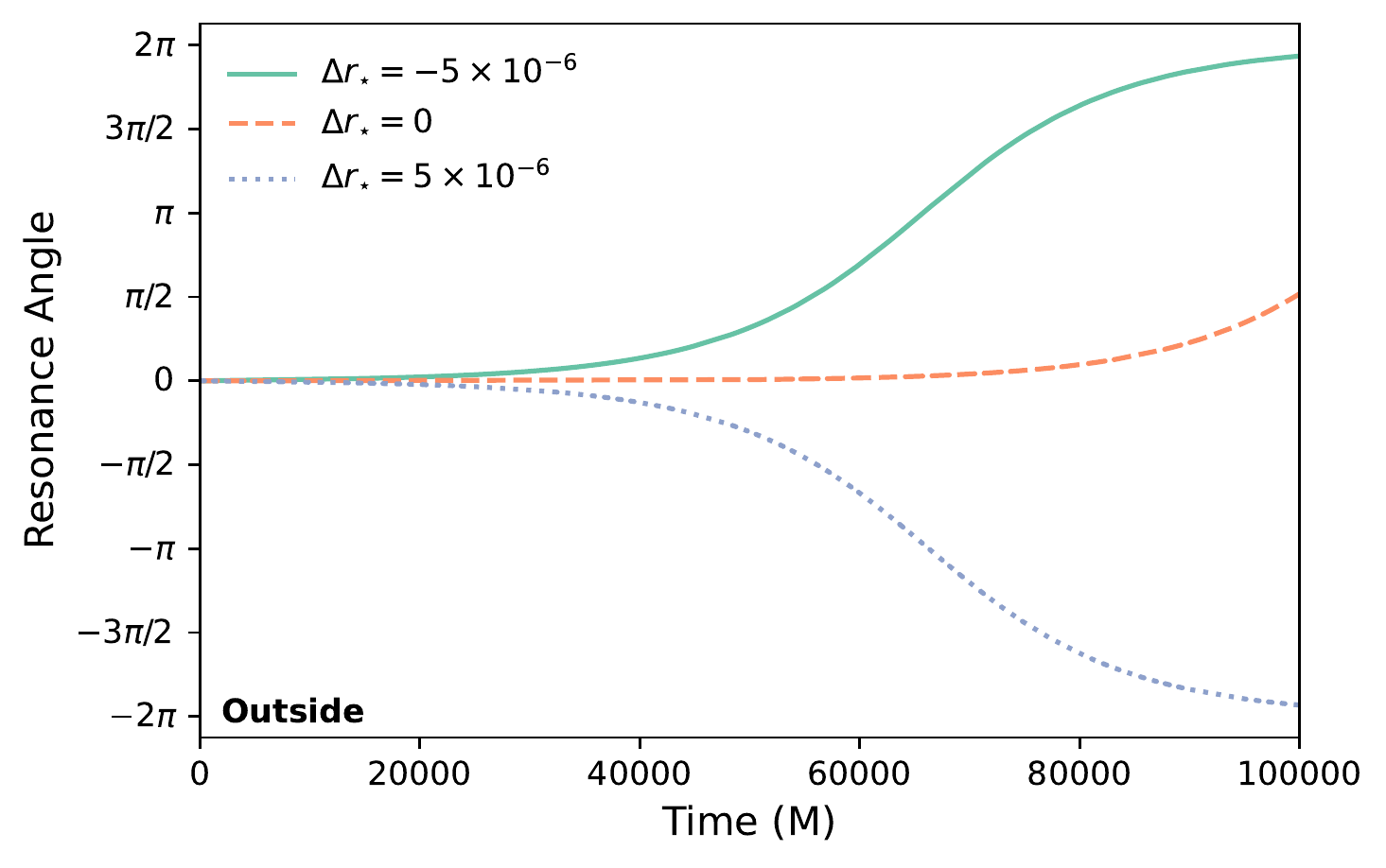}
  \end{center}
  \caption{%
    Top: Behaviour of the resonance angle inside the elliptic region. Each curve shows different values for $\Delta\psi=\psi_i - \psi_0$, where $\psi_i$ is the initial radial phase angle and $\psi_0$ is the radial phase angle of the hyperbolic fixed point. The blue dotted line is nearest to the elliptic fixed point while the green solid line is nearest to the separatrix and hyperbolic fixed point.
    Bottom: Behaviour of the resonance angle outside the elliptic region. Each curve shows different values for $\Delta r_* = r_i - r_0$, where $r_i$ is the initial radial coordinate and $r_0$ is the radial coordinate of the hyperbolic fixed point. The orange dashed line is nearest to the separatrix and the hyperbolic fixed point.}
    \label{fig:res-angle:res-sweep}
\end{figure}

To better explore the dynamics inside and outside of the elliptic
region, we further specialize to a perturbation amplitude of
$\varepsilon=10^{-7.5}$ and perform a sweep of trajectories with
initial conditions in each category. A few samples of these sweeps are
shown in Fig.~\ref{fig:res-angle:res-sweep}. From these plots, we see
that near the center of the elliptic region, trajectories become near
perfect sinusoids, where the period is approximately
amplitude-independent (the behavior of a simple harmonic oscillator).
Far from the center, near the separatrix of the region, the orbits
remain oscillatory, but with arbitrarily long periods (the behavior of
a pendulum near its separatrix).  Outside the region, we see the
resonance angle ``rotate,'' $\Psi$ monotonically decreasing or increasing
with time, depending on what side of the hyperbolic point they sit.

\section{Approximate Gravitational Waveforms}
\label{sec:gw}

Equipped with an understanding of the dynamics at resonance for this
system, we now turn to analyze the implications for detectability and
modeling. For this purpose and neglecting gravitational wave
dissipation, we construct kludge waveforms~\cite{Gair:2005ih,
  Babak:2006uv} for the perturbed system and for the unperturbed
system. We then consider the match between these to find whether or
not it is necessary to model external tidal perturbations.

With our goals set in mind, several details must be handled. First, we
must obtain the kludge waveforms from the phase space information we
have. Second, we must find a consistent way to find pairs of systems,
one perturbed and the other unperturbed, which can be identified
together and compared via waveform match calculations. And finally, we
must choose a figure of merit for estimating whether or not tidal perturbations
dominate the dynamics and to what degree.

\subsection{Construction of Kludge Waveforms}
\label{sec:gw:kludge}

We compute approximate gravitational waveforms closely following the
procedure described in Refs.~\cite{Gair:2005ih,
  Babak:2006uv,Sopuerta:2009iy}. In this approach, the motion around
the central black hole is interpreted as motion in flat spacetime, and
gravitational waves are emitted as if in flat spacetime via a
multipole decomposition~\cite{Thorne:1980ru}.

These approximate waveforms, also known as ``numerical kludges,'' are defined as $h=h_+ - i h_\times$,
where the two waveform polarizations (``plus'' and ``cross'')
are obtained from $ h_{+,\times} = \epsilon^{ij}_{+,\times} h_{ij}^{\text{TT}}/2$.
To lowest order, the metric perturbation in transverse-traceless gauge is
\begin{align}
h^{TT}_{ij}&=\left[\frac{2}{R} I^{(2)}_{ij}+\frac{4}{3R}J^{(2)}_{k(i}  \epsilon^{\,\,\,kl}_{j)}\hat n_l\right]^{\text{TT}},\label{observed-metric-pert}
\end{align}
with 
\begin{align}
    I^{(2)}_{ij}&=m\,\left[(a_ix_j+2v_iv_j + x_ia_j)\right]^{\text{STF}}\notag\\
    J^{(2)}_{ij}&=m\,\left[x_i (\vec v\times \vec a)_j
    + 2 v_i (\vec x\times \vec a)_j\right.\notag\\
    &\kern 0.7cm \left.+x_i(\vec x\times \vec j)_j+a_i(\vec x\times \vec v)_j\right]^{\text{STF}}.\label{eq:multipole-moments}
\end{align}
In these expressions, STF denotes the symmetric-tracefree projection operator, TT is the transverse-traceless projection operator, $R$ the flat-space distance from the source to the observer, and $x_i \left( t \right)$ is the set of Cartesian components of the spatial trajectory of the small compact object.
Under this approach, a trajectory is interpreted as if in flat space.

Since we work in a Newtonian analogue to Kerr spacetime, no reinterpretation is needed and we may use Eq.~\eqref{eq:coord-relations} to relate our phase-space variables to Cartesian coordinates. Furthermore, we can evaluate the velocity, acceleration, and jerk of the particle in Cartesian coordinates by evaluating $v_i=\{x_i,H\},\,a_i=\{v_i,H\},$ and $j_i=\{a_i,H\}$.

The polarization tensors $\epsilon^{ij}_{+,\times}$ are built from an orthonormal triad with two components, $p$ and $q$, chosen by the observer, and the third, $n$, in the direction of wave propagation. Explicitly, these polarization tensors are $\epsilon_{+ij}=p_i p_j - q_i q_j$ and $\epsilon_{\times ij}=2p_{\left( i \right.} q_{\left. j\right)}$. We use the most common triad, given in terms of the observation point's latitude and azimuth, $\Theta$ and $\Phi$, respectively, \cite{Babak:2006uv,Sopuerta:2009iy} given explicitly by
\be
\left\{n, p, q \right\}=\left\{\frac{\partial}{\partial r },\frac{1}{R} \frac{\partial}{\partial \Theta},\frac{1}{R \sin \Theta} \frac{\partial}{\partial \Phi}  \right\}. 
\ee
We scale the resulting strain by setting $R=1$ and choose a polar angle of $\Theta=\pi/6$ and azimuthal angle of $\Phi=\pi/4$.

Despite the approximations made during the generation of these waveforms, this procedure reproduces most of the features expected for EMRI sources and, for certain parameters, shows great agreement with more accurate Teukolsky-based waveforms~\cite{Babak:2006uv,Chua:2017ujo}. 

\subsection{Locating Comparable Trajectories}
\label{sec:gw:equiv-trajectory}

In the case of a real detection, a signal is received ($h_\text{obs}$)
and the best matching modeled signal is found ($h_\text{model}$).
Since we are not including radiation reaction, the perturbed system serves as our best proxy for the observed signal, and the unperturbed system serves as our model signal. In this manner, the two signals only differ by the tidal effect; this is loosely akin to including full general relativity in the model but neglecting the tidal effect. We therefore first find a set of initial conditions $\left\{\lambda_{\text{pert}}\right\}$ which are in the non-linear resonance for the perturbed system, and integrate them with $H$ to obtain a waveform $h_\text{obs}$.  We then seek a set of initial conditions $\left\{\lambda_{\text{unpert}}\right\}$ for the unperturbed system which, once integrated with $H_0$, produce a waveform $h_\text{model}$, which maximizes the match with $h_\text{obs}$.

To the first task, we use the analytic expressions for fundamental
frequencies found in Ref.~\cite{Eleni:2019wav} to find the radial to
polar frequency ratio associated with any set of initial
conditions. We then choose $a=0.7$, $p=1.3$, $e=0.25$, require
$f_\theta:f_r=2:3$, and then solve numerically for $\eta_{\max}\approx 0.542$. We select this $2:3$ resonance because it is expected to be the one with the strongest impact on the inspiral, i.e., producing significant phase shifts~\cite{Berry:2016bit}, it is common in several EMRI systems~\cite{Ruangsri:2013hra}, and is capable of producing sustained resonances~\cite{vandeMeent:2013sza}. 

These initial conditions place the system in
resonance and, when perturbed, produces a non-linear resonance.  With
phase angles of $\phi=0$, $\psi=\pi$, and $\chi=0$, we find
trajectories near the hyperbolic point. To find the exact location of
the hyperbolic point, we produced Poincaré sections spanning the
hyperbolic point, and assessed the necessary initial conditions by
eye.  This hyperbolic point is then used as a jumping off point to
analyzing trajectories inside the elliptic region, by adjusting the
initial radial phase angle, or outside the elliptic region, by
adjusting the turning point, $r_2$.

Ideally, finding the best matching set of unperturbed initial
conditions $\left\{\lambda_{\text{unpert}}\right\}$ would involve a full parameter search.  This is, however, quite computationally expensive, due to each evaluation requiring a full integration of initial conditions.  We therefore use the dynamical quantities of the perturbed system to make this process faster.  We obtain the fundamental frequencies of the signal by considering the evolved phase coordinates as functions of time, and
taking the slope of a linear fit.  Intuitively, this is quite similar to taking
\be
f^\text{pert}_i=\frac{\alpha^\text{pert}_{i\, \text{final}} - \alpha^\text{pert}_{i\,
    \text{initial}}}{\Delta t}
\,,
\ee
where $f^\text{pert}_i$ is the $i^\text{th}$ fundamental frequency and $\alpha^\text{pert}_i$ is the $i^\text{th}$ phase angle. We found this frequency extraction technique to perform better than a Fourier-based approach (as the one presented in, for instance, Ref.~\cite{LukesGerakopoulos:2010rc}). We then find approximately best-fitting constants of motion by requiring the fundamental frequencies $f_i^\text{unpert}$ to be the same as $f_i^\text{pert}$, and, as mentioned in Sec.~\ref{sec:unpert}, we have analytic expressions for the frequencies.\footnote{%
  Notably, frequencies in Mino time will not do here. It is possible
  to find a system which agrees in all three frequencies in Mino time
  but which has a different average value for $dt/d\lambda$ and
  therefore have different true frequencies.
}
We do not perform any search on the phase angles $(\phi, \chi, \psi)$, instead simply using the same initial phase angles.

For any waveform generated from a perturbed system, we can now
generate a nearby unperturbed waveform.  If using an unperturbed system is a good approximation, then the match between the two waveforms should be quite high (or equivalently the mismatch quite low). Given two waveforms $h_1$ and $h_2$, we use the fitting factor for match~\cite{Cutler:1994ys}:
\begin{align}
    FF(h_1, h_2):=\frac{\left<h_1 | h_2\right>}{\sqrt{\left<h_1 | h_1\right> \left<h_2 | h_2\right>}}
    \,,\label{eq:fitting-factor}
    \intertext{where the inner product is defined by}
    \left<h_1 | h_2\right>:=4R\int\frac{\Tilde{h}_1(f)\Tilde{h}_2^*(f)}{S_n(f)}df
    \,,\label{eq:inner-prod}
\end{align}
where the asterisk is a complex conjugate, $\Tilde h$ is the Fourier transform of $h$, and $S_n$ is the noise power spectral density of the detector (which we take to be the sky-averaged LISA noise).

\begin{figure}
  \begin{center}
    \includegraphics[width=0.7\linewidth]{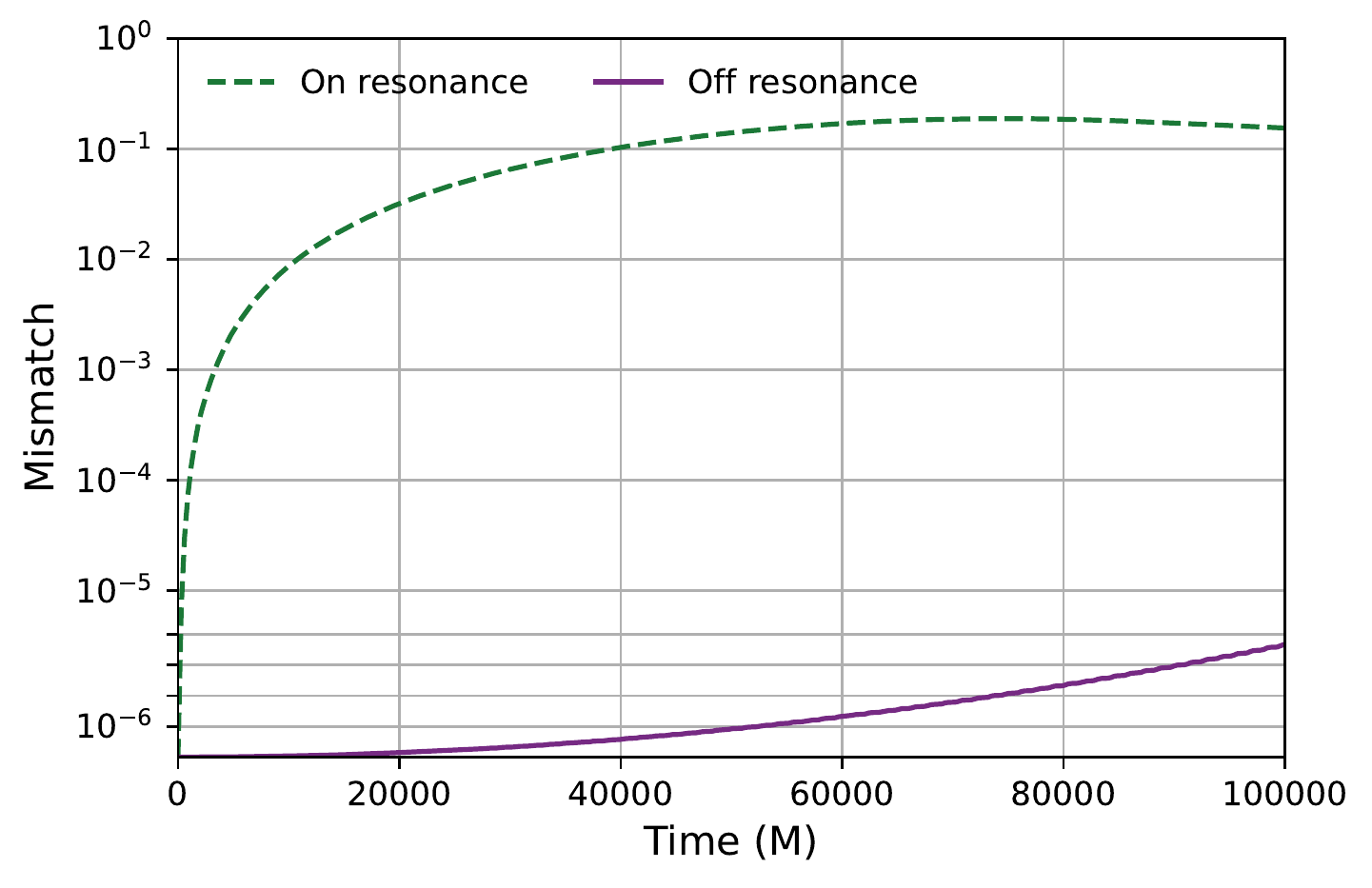}
  \end{center}
  \caption{Dephasing between perturbed and unperturbed orbits when
      in a nonlinear resonance and when far from resonance with
      $\varepsilon=10^{-8}$. For illustration
        purposes, the vertical axis has a linear scale near zero (in the finely segmented region up to $4\times 10^{-6}$) and a $\log$ scale elsewhere. Note that the off-resonance mismatch does not exceed $10^{-5}$.}
    \label{fig:res-compare}
\end{figure}

To simulate the evolving match found as observation time grows, only the first $t$ units of time of $h_1$ and $h_2$ are considered. In this way, we are able to make the above fitting factor a function of observation time. In Fig.~\ref{fig:res-compare},
we show a comparison for this mismatch when at resonance and when not at resonance. To put these values in context, $M=10^6\,M_{\odot}$ corresponds to $GM/c^{3}\sim5$ seconds, so in resonance, a perturbation of $\varepsilon=10^{-8}$ can produce a mismatch of $10^{-1}$ in about two days. Per the KAM theorem, when off-resonance, the effect of the perturbation can almost be absorbed into shifting of fundamental frequencies
(since the tori are then only deformed). This confirms our understanding that these tidal perturbations need not be modeled for EMRI systems when off-resonance, given the planned sensitivity of LISA.

\subsection{Radiation Reaction Characteristic Time}

An exact functional form of the characteristic time of radiation reaction in resonance is still not known. Assuming that the phase of the tidal force changes slowly during a resonance, the resonant self-force dephase time scales as~\cite{Flanagan:2010cd, Gupta:2022fbe}
\be
\tau_\text{rad}\sim \sqrt{\frac{4\pi}{\vec n \cdot \dot{\vec \omega}}} \sim \sqrt{\frac{1}{n\dot{\omega}_r + k\dot{\omega}_\theta}}. 
\ee
This expression, however, requires to know the rate of change of the orbital frequencies at resonance. Thus, to obtain an order of magnitude estimate, we instead choose to use a result for circular Keplerian orbits, i.e., we make the further approximation $\vec n \cdot \dot{\vec \omega}\sim \dot\omega_\text{Kepler}$. By expanding the derivative as $\dot\omega_\text{Kepler}\approx \dot H(d\omega/dH)_\text{Kepler}$, we reach to the following (rough) estimate for the characteristic time of the resonance due to radiation reaction:
\be
\tau_\text{rad}\sim \sqrt{\frac{5}{24q}}p^{15/4}|H|.\label{eq:characteristic-time}
\ee
Here we use $p$ (as defined in Eq.~\eqref{eq:turning-points}) in place of $a$, and for the initial energy we use the Hamiltonian per unit mass for the system, i.e., $E=qH$. For order of magnitude estimates, which is what we are after in this work, the most important property of this scaling is that $\tau_\text{rad} \propto 1/\sqrt{q}$.

\subsection{Dephase Time}
\label{sec:gw:dephase}

To quantify the necessity of modeling tidal effects in our system,
we will be comparing the above characteristic time to a dephase time.
For our approximate analysis, this is a much simpler calculation than
e.g.\ computing the phase jump accumulated when crossing a resonance~\cite{Gupta:2021cno}.
Instead, we use a $95\%$ match threshold to mark when the perturbed
and unperturbed systems have ``dephased''~\cite{Cutler:1994ys,
  Flanagan:1997kp,Glampedakis:2005cf}. We take this value of the
mismatch as a weak criterion, bearing in mind the limited accuracy of
our simple approximate model, whose ingredients are only qualitatively correct~\cite{Glampedakis:2005cf}.
Denote this dephase time as $\tau_\text{dephase}$.
Failing to properly model perturbations at resonance may lead
to inaccuracies in phase and detectable differences in waveforms.  Therefore we are interested in the region of parameter space where
$\tau_\text{rad} > \tau_\text{dephase}$.

\begin{figure}
  \begin{center}
    \includegraphics[width=0.7\linewidth]{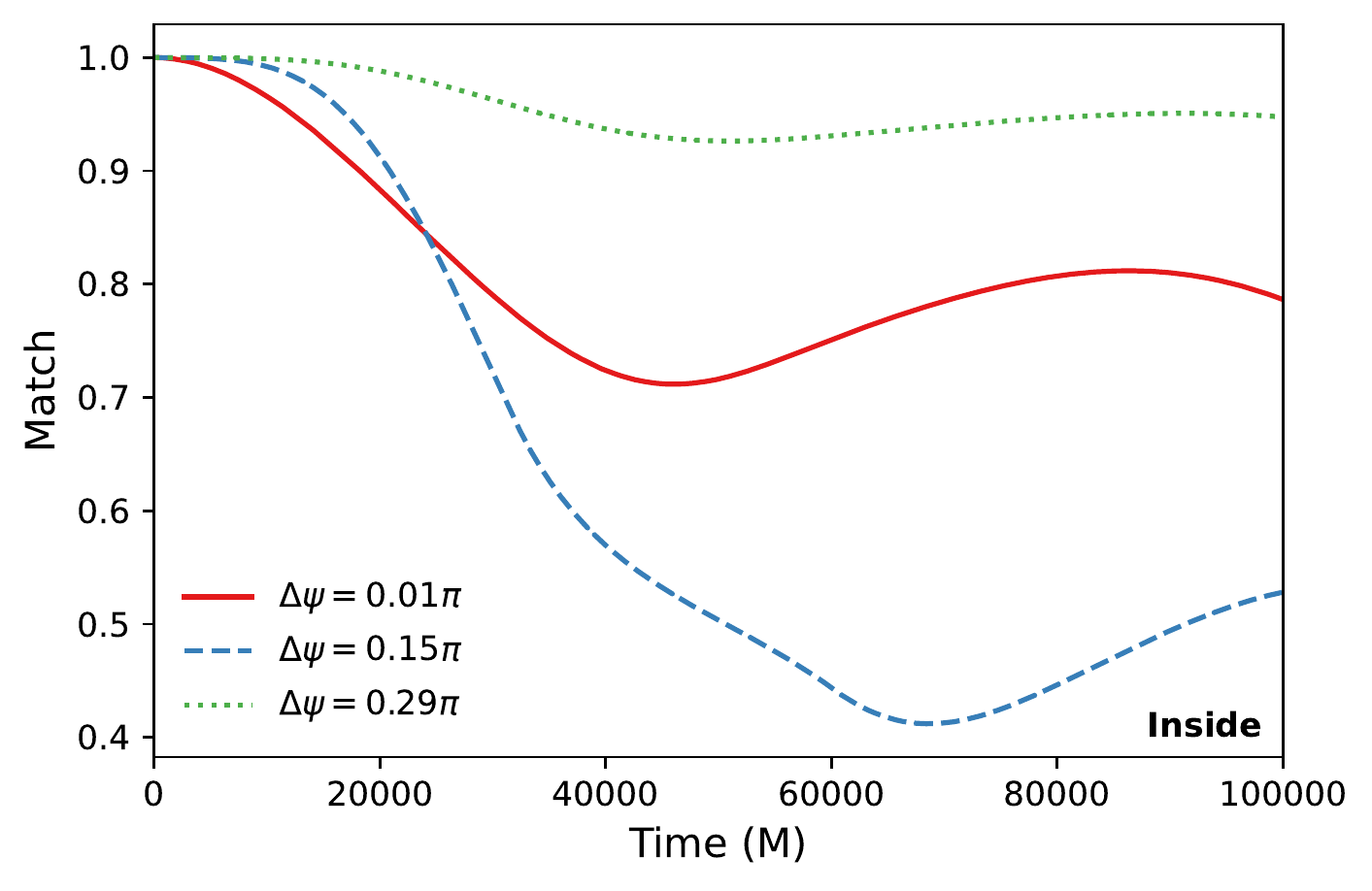}
    \includegraphics[width=0.7\linewidth]{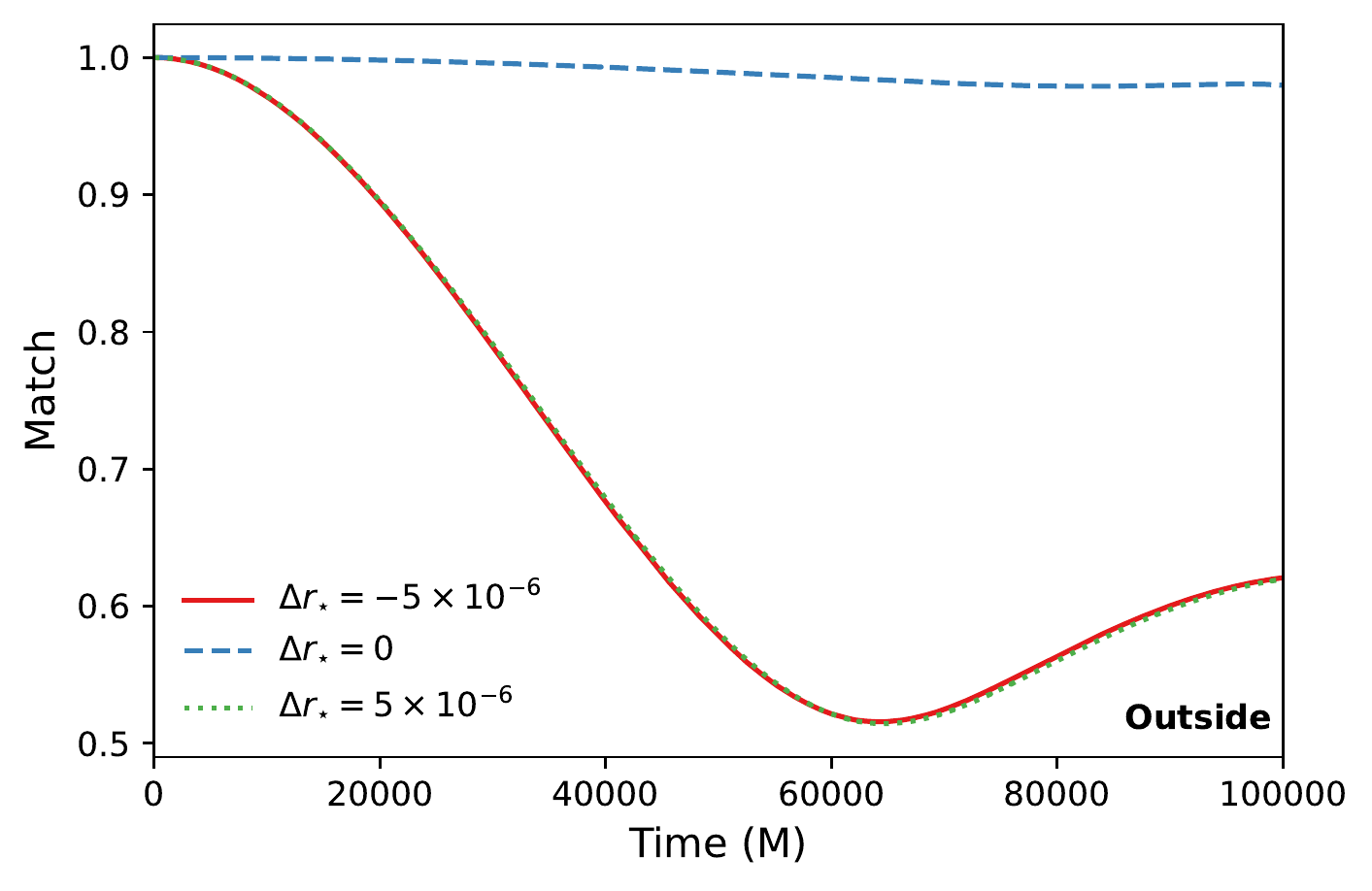}
  \end{center}
    \caption{Match comparison inside and outside of the elliptic region with a perturbation amplitude of $\varepsilon=10^{-7.5}$. We present the same values for $\Delta\psi=\psi_i - \psi_0$ and $\Delta r_* = r_i - r_0$ shown in Fig.~\ref{fig:res-angle:res-sweep}. As expected, little dephasing is seen near the elliptic fixed point (dotted green line in top figure) or near the hyperbolic fixed point (dashed blue line in bottom figure). The most representative region of initial conditions (based on several more cases not shown here) is about half way inside the elliptic region (represented by the red line in the top figure).}
    \label{fig:match-sweep}
\end{figure}

This dephase time, however, requires some further consideration. This
is because the exact time to reach the $95\%$ threshold depends on the
exact initial conditions within the resonance. This is made evident in
Fig.~\ref{fig:match-sweep}, where significantly different dephase
times can be found for the same magnitude of perturbation. Most
notably, we can find arbitrarily long dephase times very near the
hyperbolic points and near the elliptic points. We have observed
consistent dephase times roughly halfway inside the elliptic region, and so we take this to be a characteristic dephase time of the resonance.

\subsection{Comparison}
\label{sec:gw:compare}

\begin{figure}
  \begin{center}
    \includegraphics[width=0.7\linewidth]{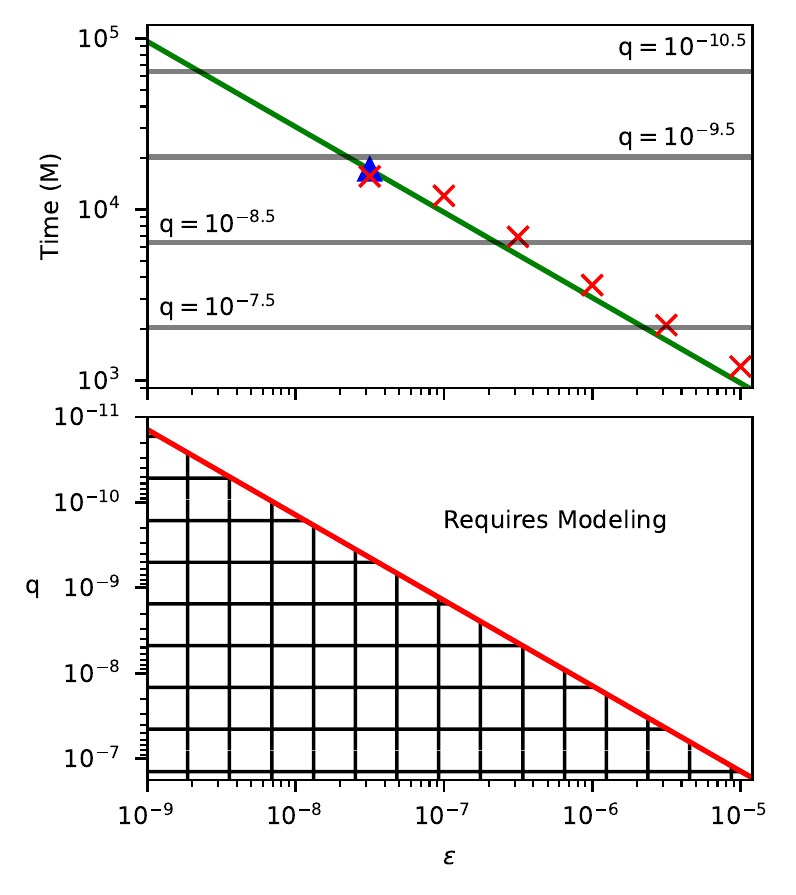}
  \end{center}
    \caption{%
    Top: Dephasing time vs. perturbation level inside the elliptic region for a 2:3 resonance. Estimated characteristic times for various mass ratios are displayed as horizontal lines.
    Red crosses are from individual simulations. The blue triangle for $\varepsilon=10^{-7.5}$ was tuned to be half way inside the elliptic region, while all other crosses were taken just outside of the elliptic region near the hyperbolic point.
    Bottom: The critical curve where the effects of resonance become dominant. A full modeling scheme will be needed for perturbations at resonance for systems above the critical curve (with larger perturbation or more extreme mass ratios).}
    \label{fig:combined_dephase_critical}
\end{figure}

In the top panel of Fig.~\ref{fig:combined_dephase_critical}, we show
dephase times found at several magnitudes of perturbation. However,
most levels of perturbation were not thoroughly analyzed throughout
the elliptic region. Most importantly, the dephase times were not
taken consistently at a point midway into the elliptic region, and so
there is more variance than would be desired. We nevertheless find a
consistent trend of dephase times following a $O (1/\sqrt
\varepsilon)$ power law, in agreement with analytical scaling arguments~\cite{MR0179413,Arnold2009,MR1656199}. Since
we are confident in the power law relation, we elect to place a fit
line based on $\varepsilon=10^{-7.5}$ only.  We performed a more
thorough search of the elliptic region at this $\varepsilon$, and it
is where we found consistent dephase times midway into the elliptic
region. 

We estimate the characteristic times for various mass ratios using
Eq.~\eqref{eq:characteristic-time} and the results are presented in the
top panel of Fig.~\ref{fig:combined_dephase_critical}.
We see that
at any given magnitude of perturbation, there is a critical mass ratio
beyond which the system will dephase due to the tidal
perturbation within the characteristic time of the resonance under radiation reaction. This region is taken
as an estimate for when tidal perturbations will need to be modeled, as shown in the bottom panel of Fig.~\ref{fig:combined_dephase_critical}. This figure shows the critical mass ratio as a function of perturbation
strength, highlighting the fact that perturbations must be modeled
for more extreme mass ratios and serving as an estimate for
where such modeling will be needed.

\section{Discussion and conclusions}
\label{sec:disc-conc}

Using a Newtonian analogue of a Kerr black hole, we performed a preliminary analysis of the effect of an external tidal field on EMRI systems.
While such effects have been previously studied, the impact tidal perturbations
have when a system is at resonance has not been thoroughly explored.
We qualitatively explored some issues in the case of perturbing the
$f_\theta:f_r=2:3$ resonance, such as the impact of the entry point
into that resonance, and we found an approximate threshold for which further
modeling will be necessary. Our results should be taken as indicative;
a genuine relativistic implementation of our method with the inclusion of gravitational self-force is needed to fully explore the interplay of tidal perturbations with self-force effects.

In our analysis, we used two measures to study this effect:
the fitting factor between waveforms from the perturbed and unperturbed systems
and the evolution of the resonance angle for the perturbed system.
Both approaches proved valuable in understanding how detection
of EMRI systems will be impacted by tidal perturbations.
The analysis of dephasing times reveals a broad region of
parameter space for which the characteristic timescale of
the tidal perturbation at resonance (the dephasing time) is
shorter than the estimated characteristic time for the analogue model.
This criteria serves as a proxy for when proper modeling of the
tidal perturbation will be necessary to accurately model the waveform.
As an example, this region of parameter space is shown in Fig.~\ref{fig:combined_dephase_critical}
for a 2:3 resonance in the analogue model, and follows the heuristic
\be \label{eq:final-param-space}
\varepsilon \gtrsim 70\, q
\,,
\ee
where $q = m/M \ll 1$ is the mass ratio
and 
$\varepsilon$ is a dimensionless scalar quantity indicating the strength of the tidal force.
For an external perturber of mass $M_*$ at a distance $d$, $\varepsilon \sim M^2 M_*/d^3$. The above scaling shows that the phase accumulated in crossing a single resonance should be negligible for small tidal perturbations, such as the astrophysically realistic regions of the parameter space.

To improve our understanding of these systems, future work is needed which simultaneously uses the Kerr spacetime, includes the gravitational self-force, and a tidal perturbation. With such a model, the region of parameter space for which tidal perturbations are important could be faithfully captured and account for more parameters, e.g., different orbital resonances (beyond the 2:3 resonance studied here), eccentricity and inclination, and alignment of the tidal force. Such work could help in constructing a robust model for traversing resonances~\cite{Gupta:2021cno}, able to account for the changes to phase space due to an arbitrary tidal force.

In addition to an estimated range for which the tidal resonance effects are relevant, we also found that different entry points into resonance can produce substantially different dynamics. This can be best seen in Figure~\ref{fig:res-angle:res-sweep}. If entering near the hyperbolic point, the resonance angle can evolve arbitrarily slowly. This is in contrast to entering between hyperbolic points where the resonance angle varies most quickly. This observation aligns with the sensitivity to initial conditions found in Ref.~\cite{Flanagan:2010cd}. A dependence between the entry point and the dynamics of the system was also reported when studying a non-Kerr solution~\cite{LukesGerakopoulos:2010rc} in the adiabatic approximation.

We found that this Newtonian analogue is a good compromise between a system complex enough to exhibit resonant effects, and still simple enough to be numerically and analytically tractable. In particular, using a Newtonian system allowed us to incorporate a tidal perturbation by simply adding a tidal potential to the Hamiltonian. In contrast, any future work in Kerr spacetime will need to utilize more sophisticated techniques, i.e., black hole perturbation theory and metric reconstruction, to include tidal effects~\cite{LeTiec:2020bos}. Our results present another example (e.g., see Refs.~\cite{Unruh:1980cg, Ray:2014sga,Torres:2016iee}) where the use of an analogue provides insight and motivates targeted studies in the full system.

The LISA mission's capacity to probe strong gravity with EMRIs depends on the ongoing endeavors to build precise GW models for these systems.  Extending the breadth of this study to the morphology and characteristic of these prolonged resonances could reveal when general resonant phenomena must be included to properly model EMRIs.
If these effects are not accounted for, they could lead to incorrect parameter estimation or fundamental biases when studying general relativity.

\ack

We would like to thank B.\ Bonga and J.\ Meyer for useful discussions and comments, and 
E.\ Flanagan and S.\ Hughes for valuable feedback on an earlier draft of this paper. A.C.-A. acknowledges funding from the Fundación Universitaria Konrad Lorenz (Project 5INV1), from Will and Kacie Snellings, support from the Simons Foundation, and thanks the Department of Physics at the University of Mississippi, where part of this work was performed, for their hospitality. Some of the computational efforts presented here were performed on the Scientific Computing Laboratory, operated and supported by Fundación Universitaria Konrad Lorenz's Engineering and Mathematics Department.
L.C.S. acknowledges D.~Nichols and K.\ Deck for helpful discussions.
The work of L.C.S. was partially supported by NSF CAREER Award
PHY--2047382.
In preparing this manuscript we made use of the \texttt{python} packages
\texttt{numpy}~\cite{harris2020array},
\texttt{scipy}~\cite{2020SciPy-NMeth},
and
\texttt{matplotlib}~\cite{Hunter:2007}.



\section*{References}
\bibliographystyle{iopart_num}
\bibliography{References}

\end{document}